\documentclass{article}
\usepackage{spconf,amsmath,graphicx,cite,citesort}
\usepackage{amsfonts}
\usepackage{pgfplots}
\usepackage{algorithm,algorithmic}
\usepackage{url}

\usepackage{bm} 

\DeclareMathOperator*{\argmin}{arg\,min}

\newcommand{\figref}[1]{Fig.~\ref{#1}}

\newcommand{\algoref}[1]{Algorithm~\ref{#1}}

\newcommand\va{\ensuremath{\mathbf{a}}}

\newcommand\vc{\ensuremath{\mathbf{c}}}
\newcommand\vd{\ensuremath{\mathbf{d}}}

\newcommand\vf{\ensuremath{\mathbf{f}}}

\newcommand\vu{\ensuremath{\mathbf{u}}}
\newcommand\vv{\ensuremath{\mathbf{v}}}

\newcommand\vx{\ensuremath{\mathbf{x}}}
\newcommand\vy{\ensuremath{\mathbf{y}}}
\newcommand\vz{\ensuremath{\mathbf{z}}}

\newcommand\valpha{\ensuremath{\bm{\alpha}}}
\newcommand\vbeta{\ensuremath{\bm{\beta}}}
\newcommand\vgamma{\ensuremath{\bm{\gamma}}}
\newcommand\vomega{\ensuremath{\bm{\omega}}}

\newcommand\mA{\ensuremath{\mathbf{A}}}

\newcommand\mC{\ensuremath{\mathbf{C}}}
\newcommand\mD{\ensuremath{\mathbf{D}}}

\newcommand\mU{\ensuremath{\mathbf{U}}}
\newcommand\mV{\ensuremath{\mathbf{V}}}

\newcommand\mPhi{\ensuremath{\bm{\Phi}}}

\title{Spectral Compressive Sensing with Polar Interpolation}
\name{Karsten Fyhn$^*$, Hamid Dadkhahi$^\dagger$, Marco F. Duarte$^\dagger$\thanks{E-mails: kfn@es.aau.dk, \{hdadkhahi,mduarte\}@ecs.umass.edu. The first two authors contributed equally to this work. This work was partially supported by The Danish Council for Strategic Research under grant number 09-067056 and by NSF under grant number ECCS-1201835.}}
\address{$^*$Dept. of Electronic Systems, Aalborg University, Denmark. \\
$^\dagger$Dept. of Electrical and Computer Engineering, University of Massachusetts Amherst, USA.}
\begin{document}

\maketitle

\begin{abstract}
Existing approaches to compressive sensing of frequency-sparse signals focuses on signal recovery rather than spectral estimation. Furthermore, the recovery performance is limited by the coherence of the required sparsity dictionaries and by the discretization of the frequency parameter space. In this paper, we introduce a greedy recovery algorithm that leverages a band-exclusion function and a polar interpolation function to address these two issues in spectral compressive sensing. Our algorithm is geared towards line spectral estimation from compressive measurements and outperforms most existing approaches in fidelity and tolerance to noise.

\end{abstract}

\section{Introduction}
\label{sec:intro}

One of the most popular thrusts in compressive sensing (CS) research has focused
on the recovery of signals that are spectrally sparse (i.e., that have a sparse
frequency-domain representation) from a reduced number of measurements~\cite{CandesRomberg2007,Tropp2010,Mishali2010,Waking2012,Duarte2012}. Such
{\em frequency-sparse signals} bring up a novel issue in the formulation of
the CS recovery problem: frequency-domain representations have a continuous parameter space, while CS is inherently rooted on discretized
signal representations.

Aiming for an increasingly dense sampling of the frequency parameter space
introduces performance issues in sparsity-leveraging algorithms. In particular,
increasing the resolution of the parameter sampling worsens the coherence of the
dictionary that provides sparsity for relevant signals. This both prevents
certain algorithms from finding the sparse representation successfully and
introduces ambiguity on the choice of representations available for a signal in
the dictionary. Initial contributions address such issues by
modifying the sparsity prior, the recovery algorithm, or both, to be tailored to
the intricacies of the signal
representation~\cite{Duarte2012,Candes2011,Tang2012,Fannjiang2012}.

Interestingly, CS recovery of frequency-sparse signals can be formalized in two
different ways: recovery of the signal samples, and recovery of the signal's
component frequencies. Previous contributions have almost exclusively
focused on the former; their performance for the latter goal is limited by
the representation leveraged during CS. Particularly, the required
discretization of the parameter space explicitly limits the performance of compressive frequency
estimation.

In this paper, we improve over existing approaches by introducing interpolation
steps within CS recovery algorithms that break the discretization barrier
implicit in CS and are able to improve the quality of frequency parameter
estimation. While such interpolation is considered briefly and integrated to a
simple recovery algorithm in~\cite{Duarte2012}, we introduce a novel polar
interpolation approach that leverages the fact that frequency-sparse signals are translation-invariant in the frequency domain. We couple polar interpolation with a
more sophisticated CS greedy recovery approach to improve the performance of
spectral CS over existing algorithms. We provide experimental evidence 
that shows improved frequency estimation performance
against approaches previously proposed for spectral CS signal recovery: in some cases, our estimates are more precise than those from the baseline
approaches, while in other cases we match the precision of the baseline with
greatly reduced computational complexity.

\section{Background and Related Work}
\label{sec:background}

Compressive sensing (CS) is a technique to simultaneously acquire and reduce the dimensionality of
sparse signals in a randomized fashion. More precisely, in the CS framework, a
signal $\vf \in \mathbb{C}^N$ is sampled by $M$ linear measurements of the form
$\vy = \mA \vf$, where $\mA$ is an $M \times N$ sensing matrix
and $M \ll N$. In practice, the measurements are acquired in the presence of noise
$\vz$, in which case we have $\vy = \mA \vf + \vz$.

In many applications, the signal $\vf$ is not sparse but has a sparse
representation in some dictionary $\mD$. In other words, we have $\vf = \mD \vx$, where
$\vx$ is $K$-sparse (i.e. $||\vx||_0 \leq K$). Under certain conditions on the
matrix $\mA$ \cite{Donoho2006,CandesRombergTao2005}, we can recover $\vx$ from the measurements $\vy$
through the following $\ell_1$-minimization problem (which we refer to as
\textit{$\ell_1$-synthesis}):
\begin{equation}
\hat{\vx} = \min_{\tilde{\vx} \in \mathbb{C}^N} ||\tilde{\vx}||_1 ~\textrm{s.t.}~ ||\mA \mD \tilde{\vx} - \vy||_2 \leq \epsilon,
\label{eq:synthesis}
\end{equation}
where $\epsilon$ is an upper bound on the noise level $||\vz||_2$. 
Note that optimal recovery of $\vx$ from the optimization in (\ref{eq:synthesis})
is feasible only when the elements of the dictionary $\mD$ form an orthonormal
basis, and thus are incoherent \cite{CandesRomberg2007,Rauhut2008}. However, in many
applications, the signal of interest is sparse in an overcomplete dictionary or a
frame, rather than in a basis.

This paper focuses on frequency-sparse signals, which can be modeled as a
superposition of $K$ complex sinusoids with arbitrary frequencies $\tilde{\vomega} = \{\omega_1, \omega_2,\ldots,\omega_K\}$.
The signal $\vf=\begin{bmatrix} f_1 & f_2 & \ldots f_N \end{bmatrix}^T$ is given
by
\begin{align}
	f_n = \sum_{k=1}^K x_k e^{j2\pi\tilde{\omega}_k n},\; \tilde{\omega}_k \in [0,1],\; n \in \{1,2,\ldots,N\}.
\end{align}
Such signals are sparse in the discrete-time Fourier transform (DTFT), when
defined using an infinite dictionary. In practice, a finite-length
representation of the signal is required, and the transform of choice is the
discrete Fourier transform (DFT). Unfortunately, the DFT coefficients for such a
frequency-sparse signal are sparse only when the frequencies of the constituent
sinusoids are integral. One way to remedy this problem would be to employ a
dictionary corresponding to a finer discretization of the Fourier
representation. We call such a dictionary a DFT frame of redundancy $c
\in \mathbb{N}$, containing $P = c\cdot N$ elements, defined as:
\begin{align} 
	\nonumber\mD &= \left[\begin{matrix} \vd(\omega_1) & \vd(\omega_2) & \cdots & \vd(\omega_P) \end{matrix}\right],\; \omega_p = \frac{p}{P}, \\
	\vd(\omega_p) &= \begin{bmatrix} d_1(\omega_p) & d_2(\omega_p) & \ldots & d_N(\omega_p) \end{bmatrix}^T, 
\end{align} 
where $d_n(\omega) = \frac{1}{\sqrt{N}}e^{j2\pi\omega n}$. However, the DFT frame violates the incoherence requirement for the dictionary \cite{Duarte2012}. 

It has recently been shown in \cite{Candes2011} that as far as the recovery of
signal $\vf$ (instead of the sparse coefficient vector $\vx$) is concerned, the coherence condition
of the dictionary is not necessary, provided that the matrix $\mD^H \mD$ is sufficiently sparse, where $(\cdot)^H$ designates the
Hermitian operation. In this case, the signal $\vf$ can be recovered via \textit{$\ell_1$-analysis}.
However, the matrix $\mD^H \mD$ is not sufficiently sparse for DFT frames.


Alternatively, one can take advantage of structured sparsity in spectral CS recovery by using a coherence inhibition model
\cite{Duarte2012}. 
The resulting structured iterative hard thresholding (SIHT) algorithm can recover the
frequency-sparse signal with a DFT frame by avoiding dictionary elements with high coherence. 
A variation of this method uses
a band-exclusion function to achieve the same avoidance~\cite{Fannjiang2012}.
We can define the $\eta$-coherence band of the index set $S$ as
\begin{align}
	B_\eta(S) &= \bigcup_{k\in S} \{i \:|\: \mu(i,k) > \eta\},\; i \in \{1,2,\ldots,P\},
	\label{eq:bomp}
\end{align}
where $\mu(i,k)=|\langle\vd(\omega_i),\vd(\omega_k)\rangle|$ is the coherence between two atoms in the dictionary. The authors use the band-exclusion function to avoid selecting coherent dictionary elements in various greedy algorithms, including Band-excluded Orthogonal Matching Pursuit (BOMP).

More recently, it has been shown that one can recover a frequency-sparse signal
from a random subset of its samples using atomic norm minimization \cite{Tang2012}.
The atomic norm of $\vf$ is defined as the size of the smallest scaled convex hull
of a continuous dictionary of complex exponentials. 
Thus, the recovery procedure searches over a continuous dictionary rather than a
discretized one. The atomic norm minimization can be implemented as a
semidefinite program (SDP), which can be computationally expensive. 
In addition, this formulation does not account for measurement noise, and it is not clear if guarantees can be given for arbitrary measurement settings. 
Nonetheless, \cite{Tang2012} motivates our formulation of algorithms that push past the
discretization of the frequency parameter space.

\section{Polar Interpolation \\ for Frequency Estimation}
\label{sec:print}
One way to remedy the discretization of the frequency parameter space implicit in CS is to use interpolation.
In \cite{Ekanadham2011}, a \textit{polar interpolation} approach for
translation-invariant signals has been derived. Such signals
can be written as a linear combination of shifted versions of a waveform. In a
nutshell, the interpolation procedure exploits the fact that translated versions of a
waveform form a manifold which lies on the surface of a hypersphere. Thus, any
sufficiently small segment of the manifold can be well-approximated by an arc of
a circle, and an arbitrarily-shifted waveform can be closely approximated by a point 
in such arc.

The complex exponentials that compose a DFT frame also form a
manifold over a hypersphere, and thus can be approximated by an arc of a circle.
This is motivated by the fact that complex exponentials have
translation-invariant Fourier transforms, which correspond to an isometric rotation 
of the time-domain vectors.
In this case, the DFT frame samples the frequency parameter space with a steps size $\Delta = 1/c$, and we approximate a segment
of the manifold $\vd(\tilde{\omega}_i) : \tilde{\omega}_i \in
[\omega_p-\tfrac{\Delta}{2} ,\omega_p+\tfrac{\Delta}{2}]$ by a circular arc
containing the three exponentials $\{\vd(\omega_p-\tfrac{\Delta}{2}),
\vd(\omega_p), \vd(\omega_p+\tfrac{\Delta}{2})\}$.
Making use of trigonometric identities, the polar interpolator approximates exponentials 
$\vd(\tilde{\omega}_i)$, $\tilde{\omega}_i \in [\omega_p-\tfrac{\Delta}{2},\omega_p+\tfrac{\Delta}{2}]$, 
using linear combinations of the three exponentials \cite{Ekanadham2011}:
\begin{align*}
\nonumber \vd(\tilde{\omega}_i)  \approx \vc(\omega_p) + r \cos \left(\frac{2 \tilde{\omega}}{\Delta} \theta \right) \vu(\omega_p) + r \sin \left(\frac{2 \tilde{\omega}}{\Delta} \theta \right) \vv(\omega_q), \\
\begin{bmatrix} \vc(\omega_p)^T \\ \vu(\omega_p)^T \\ \vv(\omega_p)^T \end{bmatrix} = \begin{bmatrix} 1 & r\cos(\theta) & -r\sin(\theta) \\ 1 & r & 0 \\ 1 & r\cos(\theta) & r\sin(\theta) \end{bmatrix}^{-1} \begin{bmatrix} \vd(\omega_q-\tfrac{\Delta}{2})^T \\ \vd(\omega_p)^T \\ \vd(\omega_p+\tfrac{\Delta}{2})^T \end{bmatrix},
\end{align*}
where $r$ is the $\ell_2$ norm of each element of the dictionary and $\theta$ is the
angle between $\vd(\omega_p)$ and $\vd(\omega_p - \tfrac{\Delta}{2})$. 
In order to extend the above approximation to sums of $J$ exponentials with frequencies $\Omega = \{\omega_1, \omega_2, \ldots, \omega_J\}$, we define:
\begin{align}
	\tilde{\vf} &= \mC(\Omega)\valpha - \mU(\Omega)\vbeta - \mV(\Omega)\vgamma,
\end{align}
\begin{align}
	\mC(\Omega) &= \begin{bmatrix} \vc(\omega_1) & \vc(\omega_2) & \cdots & \vc(\omega_J)\end{bmatrix}, \nonumber \\
	\label{eq:f_approx}\mU(\Omega) &= \begin{bmatrix} \vu(\omega_1) & \vu(\omega_2) & \cdots & \vu(\omega_J)\end{bmatrix}, \\
	\mV(\Omega) &= \begin{bmatrix} \vv(\omega_1) & \vv(\omega_2) & \cdots & \vv(\omega_J)\end{bmatrix}, \nonumber
\end{align}
where  $\valpha$ represents the amplitude of the signal and $\vbeta$ and
$\vgamma$ controls the frequency translations. The three coefficient vectors
can be estimated using the following constrained convex optimization
problem \cite{Ekanadham2011}:
\begin{align}
	\label{eq:convex_optimization}
    (\valpha,\vbeta,\vgamma) &= \mathrm{T}(\vy, \mA, \Omega) \\
    \nonumber&=\argmin_{\valpha,\vbeta,\vgamma} \frac{1}{2\sigma^2} || \vy - \mA\tilde{\vf}||_2^2 + ||\valpha||_1 \\
    \nonumber&\text{s.t. } \left\{\begin{matrix} \alpha_j \geq 0, \\ \sqrt{\beta_j^2 + \gamma_j^2} \leq \alpha_j^2 r^2, \\ \alpha_j r\cos(\theta) \leq \beta_j \leq \alpha_j r, \end{matrix}\right\} \;\text{for } j=1,\ldots,J,
\end{align}
where $\mA$ is the measurement matrix, and $\vy$ is the received compressed
signal. The constraints for the optimization problem ensure that the solution consists
of points on the arcs used for approximation. The first constraint ensures we have
only nonnegative signal amplitudes. The second enforces the trigonometric relationship
among each triplet $\alpha_j$, $\beta_j$, and $\gamma_j$. The last constraint
ensures that the angle between the solution and $\vd(\omega_j)$ is restricted to
the interval $[0,\theta]$.
It is necessary to scale $\vbeta$ and $\vgamma$ after the optimization problem~\cite{Ekanadham2011}:
\begin{align}
	(\beta_j, \gamma_j) \leftarrow \left(\frac{\beta_j\alpha_j r}{\sqrt{\beta_j^2+\gamma_j^2}}, \frac{\gamma_j\alpha_j r}{\sqrt{\beta_j^2+\gamma_j^2}}\right).
\end{align}
This is because the inequality of the second constraint should in fact be an equality. However, the equality would violate the convexity assumption of the optimization.
After this normalization, we obtain the signal estimate from (\ref{eq:f_approx}) and the 
frequency estimates using the one-to-one relation 
\begin{align} 
	\label{eq:polarint}\alpha_j \vc(\omega_j) + \beta_j \vu(\omega_j) + \gamma_j \vv(\omega_j) = \alpha_j \vd\left(\omega_j + \tfrac{\Delta}{2\theta}\tan^{-1}(\tfrac{\gamma_j}{\beta_j})\right).
\end{align}
The optimization (\ref{eq:convex_optimization}), when applied with all parameter values used in the dictionary $D$, is named continuous basis pursuit (CBP) in \cite{Ekanadham2011}:
\begin{align}
	(\valpha,\vbeta,\vgamma) &= \mathrm{T}(\vy, \mA, \Omega_{CBP}),
	\label{eq:cbp}
\end{align}
where $\Omega_{CBP} = \{\omega_1, \omega_2, \ldots, \omega_P\}$ is the set of all frequencies that appear in the DFT frame for our application of interest.
As posed, CBP has a high computational complexity: it operates on matrices
of size $3N$, whereas other CS algorithms operate on matrices of size $N$.
However, its interpolation step has one important advantage: translation-invariance and interpolation enables CBP to reconstruct arbitrary frequency-sparse signal while requiring only a small subset of the corresponding dictionary.
This makes it possible to incorporate the convex optimization solver
into a greedy algorithm that quickly finds a rough estimate, which is then
improved upon by a convex optimization solver.

\section{Band-Excluded \\ Interpolating Subspace Pursuit}
\vskip-0.2cm
We incorporate the convex optimization (\ref{eq:convex_optimization}) and
band-exclusion (\ref{eq:bomp}) in a Subspace Pursuit algorithm\cite{Dai2009}. We
call this algorithm Band-Excluded Interpolating Subspace Pursuit (BISP), which
is shown in \algoref{algo:bisp}.

In the algorithm initialization, the best $K$ correlating atoms are found and
stored in $S^n$ by generating a proxy for the sparse signal. The $K$ atoms are
found iteratively, which deviates from the original Subspace Pursuit algorithm
where the $K$ atoms are found in one step. In each iteration, we trim the proxy
based on the found atom and the band exclusion function $B_\eta(S)$, as defined
in (\ref{eq:bomp}).
In the main loop, we find the $K$ best atom indices and add them to $S^n$.
From $S^n$, we form a set $\Omega$ consisting of all frequencies corresponding
to the indices in $S^n$ along with all adjacent indices. This is necessary
because the frequencies present in $\vy$ may not be sufficiently incoherent and
may therefore skew the peaks of the proxy estimate. Therefore, as a precaution,
we include the closest neighbors on each side. The set $\Omega$ is input to the
convex optimization in (\ref{eq:convex_optimization}) along with the measurement
matrix and the received signal.

In practice, we found that for noisy measurements it is often preferable to
move the minimization objective $||\vy - \mA\tilde{\vf}||_2^2$ in
(\ref{eq:convex_optimization}) into a constraint. Moving this fidelity measure from
the objective function to a constraint causes the optimization to return the
sparsest set of coefficients that yields measurements within the noise range of
the observation. If the output is non-existent or trivial, we move the fidelity
metric from the objective function to the constraint (or vice versa).
\begin{algorithm}[h]
\caption{BISP}
\label{algo:bisp}
\begin{algorithmic}
	\STATE \textbf{INPUTS:} Compressed signal $\vy$, sparsity $K$, measurement matrix $\mA$ and spacing between dictionary elements $\Delta$.
	\STATE \textbf{OUTPUTS:} Reconstructed signal $\tilde{\vf}$ and frequency estimates $\tilde{\vomega}$.
	\STATE \textbf{INITIALIZE:} $\mPhi=\mA\mD$, $i=1, S^0 = \emptyset$
	\WHILE{$i \leq K$} 
		\STATE $S^0 = S^0 \cup \arg\max_i |\langle\vy,\mPhi_i\rangle|,\:i \not\in B_0(S^0),\; i = i + 1$
	\ENDWHILE
	\STATE $\vy_r^0=\vy-\mPhi_{S^0}\mPhi_{S^0}^\dagger\vy$,\; $n=1$
	\STATE \textbf{LOOP:}
	\REPEAT 
		\STATE $i=1, S^{n} = S^{n-1}$
		\WHILE{$i \leq K$} 
			\STATE $S^n = S^n \cup \arg\max_i |\langle\vy,\mPhi_i\rangle|,\:i \not\in B_0(S^n),\; i = i + 1$
		\ENDWHILE		
		\STATE $\va = (\mPhi_{S^n})^\dagger \vy$
		\STATE $S^n = \text{supp}(\text{thresh}(\va,K))$
		\STATE $\Omega = \cup\{\Delta(s-1),\Delta s,\Delta(s+1) | s \in S^n\}$
		\STATE From $\mathrm{T}(\vy, \mA, \Omega)$ obtain $\tilde{\vf}$ and $\tilde{\vomega}$ using (\ref{eq:polarint}) and (\ref{eq:f_approx})
		\STATE $\vy_r^n=\vy-\mA\tilde{\vf},\; n = n + 1$	
	\UNTIL{$||\vy_r^n||_2 > ||\vy_r^{n-1}||_2 \vee n \leq K$}
\end{algorithmic}
\end{algorithm} 

\vskip-0.2cm
\section{Numerical Experiments}
\vskip-0.25cm
To evaluate \algoref{algo:bisp}, we have performed two numerical experiments.\footnote{The documentation and code for these experiments are made
freely available at \url{http://www.sparsesampling.com/scspi}, following the principle of
Reproducible Research \cite{Vandewalle2009}.} 
We generated frequency-sparse signals of length $N=100$ containing $K=4$ complex
sinusoids with frequencies selected uniformly at random. We used a DFT frame with
$c=5$ ($\Delta=0.2$Hz), and considered well-separated tones so that no two tones
are closer than $1$Hz of each other. We performed Monte Carlo experiments and
averaged over $30$ experiments. As measurement matrix\footnote{For the SDP
algorithm we used a random subsampling matrix, as the algorithm is only defined
for such a measurement matrix. The authors would like to thank Gongguo Tang for
providing the implementation of SDP.
} we used a Gaussian matrix $\mA\in\mathbb{R}^{M\times N}$. We set $M=\kappa N$,
where $\kappa\in(0,1]$ is the CS subsampling rate. We compare our proposed
\algoref{algo:bisp} with six state-of-the-art methods:
$\ell_1$-synthesis, $\ell_1$-analysis, SIHT, SDP, BOMP, and CBP. As performance
measure, we use the Hungarian algorithm \cite{Kuhn1955,Munkres1957} to find the best matching between the
estimated and true frequencies. For the algorithms that return a dense 
DFT coefficient vector or a reconstructed signal ($\ell_1$-synthesis,
$\ell_1$-analysis, SIHT, and SDP), we apply the MUSIC algorithm
\cite{Stoica1997} on the reconstructed signal to estimate its frequencies. In
the BISP and BOMP algorithms, we exclude atoms with coherence $\eta > 0.25$ using (\ref{eq:bomp}).

\begin{figure}[t] \centering \newlength\figureheight \newlength\figurewidth
\setlength\figureheight{4.2cm} \setlength\figurewidth{0.4\textwidth}
%
%
%
%

\definecolor{mycolor1}{rgb}{0,1,1}
\definecolor{mycolor2}{rgb}{1,0,1}
\definecolor{mycolor3}{rgb}{1,1,0}

\begin{tikzpicture}

\begin{semilogyaxis}[%
view={0}{90},
width=\figurewidth,
height=\figureheight,
scale only axis,
xmin=10, xmax=100,
xlabel={Number of measurements (M)},
ymin=1e-07, ymax=100,
yminorticks=true,
ylabel={Average cost in frequency estimation}]
\addplot [
color=blue,
solid,
line width=2.0pt,
forget plot
]
coordinates{
 (10,58.2991880713257)(20,11.6350326033208)(30,0.215629729040323)(40,0.133546496427029)(50,0.0985615715409764)(60,0.0638300753205191)(70,0.0425540175902785)(80,0.0338443422284922)(90,0.0208945514220533)(100,2.60647171993848e-07) 
};
\addplot [
color=red,
solid,
line width=2.0pt,
forget plot
]
coordinates{
 (10,49.1358928861996)(20,0.233983275015422)(30,0.00700196814350685)(40,0.00319850491479686)(50,0.00156445813985269)(60,0.000929284788932357)(70,0.000527894780519482)(80,0.000317865280528684)(90,0.000179880940875125)(100,2.59436356545943e-07) 
};
\addplot [
color=mycolor1,
solid,
line width=2.0pt,
forget plot
]
coordinates{
 (10,58.0159664983716)(20,48.1904789741925)(30,8.96323501929013)(40,17.9180379924711)(50,14.832698029791)(60,12.7843908144569)(70,16.4241321031342)(80,5.64577392864539)(90,13.9753180578247)(100,19.9054646944785) 
};
\addplot [
color=green,
solid,
line width=2.0pt,
forget plot
]
coordinates{
 (10,54.9651874217174)(20,0.00064094103005754)(30,2.44376241376187e-07)(40,2.57472028704816e-07)(50,2.65043790603769e-07)(60,2.51910519866172e-07)(70,2.92966639311073e-07)(80,2.86634039992073e-07)(90,2.59491958668849e-07)(100,2.89390763027105e-07) 
};
\addplot [
color=mycolor2,
solid,
line width=2.0pt,
forget plot
]
coordinates{
 (10,68.1176577239224)(20,5.78453662519244)(30,0.362978225426276)(40,0.35928326527865)(50,0.275849334971289)(60,0.285901257265006)(70,0.24548073743415)(80,0.261824982322361)(90,0.257081779022895)(100,0.219366955691655) 
};
\addplot [
color=black,
solid,
line width=2.0pt,
forget plot
]
coordinates{
 (10,30.4502607345078)(20,0.00187571559305261)(30,0.00170307320725981)(40,0.0018203879771145)(50,0.00147316684558186)(60,0.00154135226784303)(70,0.00149781741058695)(80,0.00150767194057914)(90,0.0011180728003643)(100,0.00134497424127657) 
};
\addplot [
color=mycolor3,
solid,
line width=2.0pt,
forget plot
]
coordinates{
 (10,56.4958609696249)(20,3.23878329212857)(30,0.526330058984779)(40,0.00275075447898851)(50,0.00230369146740463)(60,0.00234115188856011)(70,0.00227241680681622)(80,0.00226501861411703)(90,0.00225347774590858)(100,0.00237922198626252) 
};
\end{semilogyaxis}
\end{tikzpicture}%
	\caption{Frequency estimation performance in noise-less case. The legend is shown in \figref{fig:Fig2}.} 
	\label{fig:Fig1}
\end{figure}
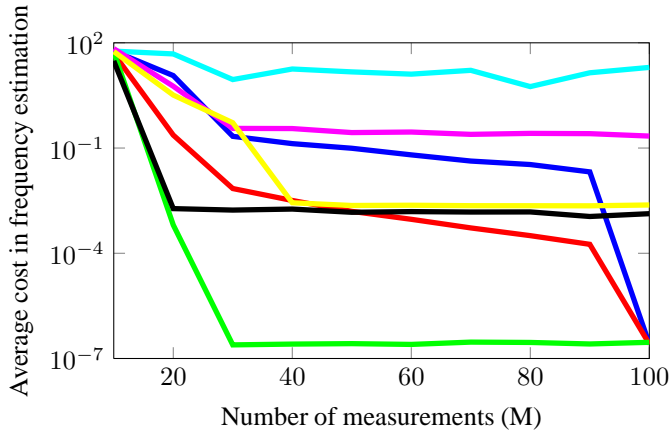

For the first experiment, we explore a range of subsampling ratios $\kappa$ with
noiseless measurements to verify the level of compression that allows for
successful estimation. We set $\epsilon=10^{-10}$ for the relevant algorithms.
The result of the numerical experiment is shown in Figure \ref{fig:Fig1}.
In the noiseless case, SDP obtains the best result. The polar interpolation
algorithms (CBP and BISP) both converge to a given estimation precision, which
corresponds to the level of approximation error. When the number of measurements
$M$ is sufficiently small, CBP outperforms $\ell_1$-synthesis. The performance of
BOMP and SIHT is worst among the algorithms tested. Surprisingly, while the DFT
coefficients $\vx$ found by $\ell_1$-synthesis are not sparse and do not match the original 
frequencies, the signal $\vf$ is still reconstructed accurately, and so the MUSIC algorithm 
recovers the frequencies adequately.

For the second experiment, we include measurement noise in the signal model.
We fix $\kappa=0.5$ and vary the signal-to-noise ratio (SNR) from $0$ to
$20$~dB. 
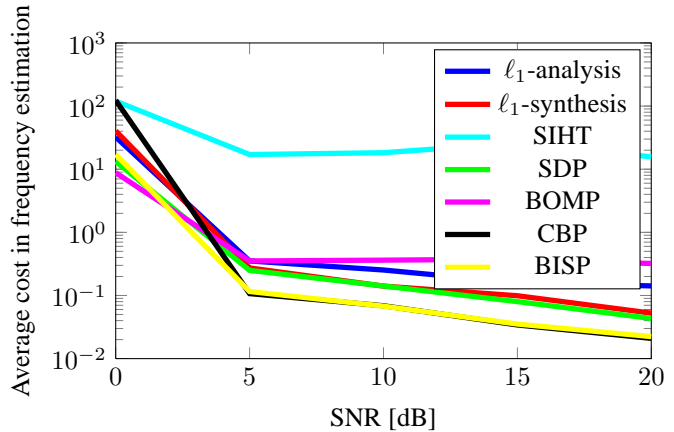
\begin{figure}[t]
	\centering
	\setlength\figureheight{4.2cm} 
	\setlength\figurewidth{0.4\textwidth}
%
%
%
%

\definecolor{mycolor1}{rgb}{0,1,1}
\definecolor{mycolor2}{rgb}{1,0,1}
\definecolor{mycolor3}{rgb}{1,1,0}

\begin{tikzpicture}

\begin{semilogyaxis}[%
view={0}{90},
width=\figurewidth,
height=\figureheight,
scale only axis,
xmin=0, xmax=20,
xlabel={SNR [dB]},
ymin=0.01, ymax=1000,
yminorticks=true,
ylabel={Average cost in frequency estimation},
legend style={align=left}]
\addplot [
color=blue,
solid,
line width=2.0pt
]
coordinates{
 (0,32.3342859558113)(5,0.352430879645507)(10,0.253260336609565)(15,0.159858364394002)(20,0.141779447380089) 
};
\addlegendentry{$\ell_1$-analysis};

\addplot [
color=red,
solid,
line width=2.0pt
]
coordinates{
 (0,40.5256642288224)(5,0.273145279729115)(10,0.140833671334637)(15,0.0990916038411578)(20,0.0525943507829534) 
};
\addlegendentry{$\ell_1$-synthesis};

\addplot [
color=mycolor1,
solid,
line width=2.0pt
]
coordinates{
 (0,119.783532331962)(5,17.0853736401404)(10,18.2478454630452)(15,26.0557911875144)(20,15.7109286608097) 
};
\addlegendentry{SIHT};

\addplot [
color=green,
solid,
line width=2.0pt
]
coordinates{
 (0,13.2697231430714)(5,0.250028444944173)(10,0.141286431716162)(15,0.0800015066374043)(20,0.0432950332052117) 
};
\addlegendentry{SDP};

\addplot [
color=mycolor2,
solid,
line width=2.0pt
]
coordinates{
 (0,8.81185028745724)(5,0.35262285253621)(10,0.362680690221498)(15,0.376850089010246)(20,0.321133222878509) 
};
\addlegendentry{BOMP};

\addplot [
color=black,
solid,
line width=2.0pt
]
coordinates{
 (0,123.674297595295)(5,0.107133559595353)(10,0.0687329266977532)(15,0.0341612904040909)(20,0.0209508268095021) 
};
\addlegendentry{CBP};

\addplot [
color=mycolor3,
solid,
line width=2.0pt
]
coordinates{
 (0,17.1235404952769)(5,0.115604919808397)(10,0.0678031675869975)(15,0.0352704253072041)(20,0.0223111081392568) 
};
\addlegendentry{BISP};

\end{semilogyaxis}
\end{tikzpicture}%
	\caption{Frequency estimation performance in noisy case.} 
	\label{fig:Fig2}
\end{figure}
In the noisy case, the polar interpolation algorithms perform best. This is
because their interpolation step relies less on the sparsity of the signal and
more on the known signal model and the fitting to a circle on the manifold. 
Additionally, the presence of noise renders the measurements non-sparse in the
dictionaries used by the non-interpolating algorithms, hindering their performance.

The computation time of the algorithms is also of importance, and we have listed
the average computation times in Table \ref{tab:comp_times}.
\begin{table}[t!]
\centering
\small
\begin{tabular}{|l|l|l|}\hline
						& Noiseless & Noisy \\\hline
	$\ell_1$-analysis 	& 9.5245	& 8.8222\\\hline
	$\ell_1$-synthesis 	& 2.9082 	& 2.7340\\\hline
	SIHT 				& 0.2628 	& 0.1499\\\hline
	SDP 				& 8.2355 	& 9.9796\\\hline
	BOMP 				& 0.0141 	& 0.0101\\\hline
	CBP 				& 46.9645 	& 40.3477\\\hline
	BISP 				& 5.4265 	& 1.4060\\\hline
\end{tabular}
\caption{Average computation times in seconds.}
\label{tab:comp_times}
\end{table}
We observed that most algorithms exhibit computation time roughly independent of
$M$, with the exception of $\ell_1$-synthesis and CBP\footnote{See results at \url{http://www.sparsesampling.com/scspi}.}. The table shows that
the excellent performance of SDP in Figure \ref{fig:Fig1} is tempered by its high
computational complexity, as well as its lack of flexibility on the measurement scheme. Moreover, the relaxation in BISP that accounts for
the presence of noise reduces its computation time, increasing its performance
advantage over SDP and CBP.

%





\bibliographystyle{IEEEbib}
\bibliography{references}

\end{document}